\documentclass{emulateapj}

\def\Q{\ifmmode\mathcal{Q}\else$\mathcal{Q}$\fi}
\newcommand{\lmst}{$\ell_{\rm MST}$}

\newcommand{\lsim}{\ \raise
-2.truept\hbox{\rlap{\hbox{$\sim$}}\raise5.truept\hbox{$<$}\ }}
\newcommand{\gsim}{\ \raise
-2.truept\hbox{\rlap{\hbox{$\sim$}}\raise5.truept\hbox{$>$}\ }}

\slugcomment{accepted by ApJ}

\shorttitle{Clustering Behavior of PMS stars in NGC 346}
\shortauthors{Schmeja, Gouliermis \& Klessen}

\begin{document}

\title{The Clustering Behavior of Pre-Main Sequence Stars in NGC~346 in the Small Magellanic Cloud}

\author{Stefan Schmeja}
\affil{Zentrum f\"ur Astronomie der Universit\"at Heidelberg, 
              Institut f\"ur Theoretische Astrophysik, Albert-Ueberle-Str.~2, 
                  69120 Heidelberg, Germany}
\email{sschmeja@ita.uni-heidelberg.de}

\author{Dimitrios A. Gouliermis}
\affil{Max-Planck-Institut f\"ur Astronomie, K\"onigstuhl 17, 
              69117 Heidelberg, Germany}
\email{dgoulier@mpia.de}

\and

\author{Ralf S. Klessen}
\affil{Zentrum f\"ur Astronomie der Universit\"at Heidelberg, 
              Institut f\"ur Theoretische Astrophysik, Albert-Ueberle-Str.~2, 
                  69120 Heidelberg, Germany}
\email{rklessen@ita.uni-heidelberg.de}

\begin{abstract}
We present evidence that the star-forming region NGC~346/N66 in the Small
Magellanic Cloud is the product of hierarchical star formation, probably
from more than one star formation event. We investigate the spatial
distribution and clustering behavior of the pre-main sequence (PMS)
stellar population in the region, using data obtained with {\sl Hubble
Space Telescope}'s Advanced Camera for Surveys. By applying the 
nearest neighbor and minimum spanning tree methods on the rich
sample of PMS stars previously discovered in the region we identify ten
individual PMS clusters in the area and quantify their structures. The
clusters show a wide range of morphologies from hierarchical multi-peak
configurations to centrally condensed clusters. However, only about 40 per
cent of the PMS stars belong to the identified clusters. The central
association NGC~346 is identified as the largest stellar concentration,
which cannot be resolved into subclusters. Several PMS clusters are
aligned along filaments of higher stellar density pointing away from the
central part of the region. The PMS density peaks in the association
coincide with the peaks of [{\ion{O}{3}}] and  8~$\mu$m emission. While more
massive stars seem to be concentrated in the central association when
considering the entire area, we find no evidence for mass segregation
within the system itself.
\end{abstract}


\section{Introduction}

Star formation is generally thought to be clumped into a hierarchy of
structures, from small multiple systems to giant stellar complexes and
beyond \citep[e.g.][]{ef+el98}, and the interaction
between gravity and supersonic turbulent motions has been proposed as a
major player in the formation of hierarchical structures 
\citep{maclow+klessen04,elm06,ballesteros07,mckee+ostriker07}. 
The interstellar medium has also been observed
to be arranged in a hierarchical structure \citep{scalo85,vazquez04}, 
sometimes described as fractal,
from the scales of the largest giant molecular clouds (GMCs) down to
individual cores and embedded clusters, which are sometimes hierarchical
themselves. 

The velocity structure in such turbulent regions is suggested to be
scale-free, meaning that the same physical processes and the same
velocity-separation relations occur over a wide range of absolute scales
\citep[e.g.][]{elm06}. According to this hypothesis, large
clumps originate from the large velocities that compress the gas on
large scales, and small clumps result from the compression of the gas
characterized by small velocities on small scales. Considering that
stars are formed by the collapse of such hierarchical structured clumps,
they are themselves hierarchically structured. Indeed, stars are mostly
formed in groups \citep[e.g.][]{sta+pal05}, which can be
seen as the bottom parts of this hierarchy \citep[e.g.][]{elm00}. 

Naturally, the structure of such systems, i.e.\ the spatial distribution
of their members, may hold important information on the mechanism and
the initial conditions of their formation. However, a thorough
investigation of the structural behavior of star formation in
environments different than our Galaxy was not possible, until recently,
due to observational limitations. In this paper we present such an
investigation for the brightest star-forming region in the Small
Magellanic Cloud (SMC).

\begin{figure}
\epsscale{1.1}
\plotone{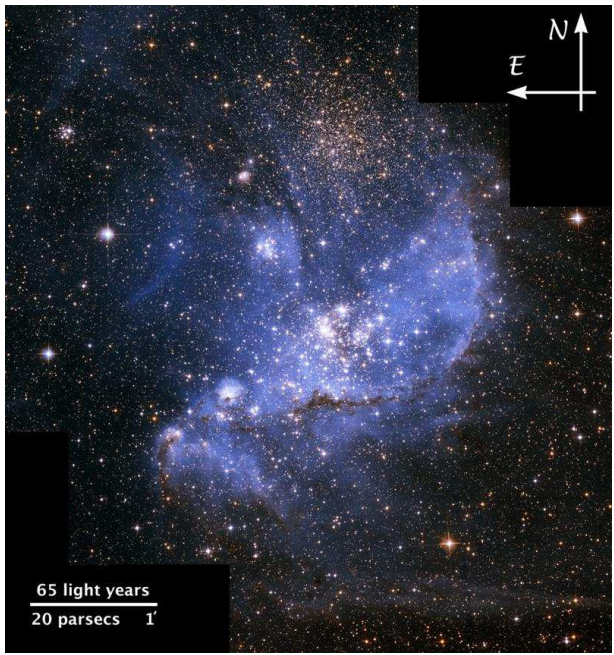}
\caption{Color-composite image of the three partially overlapping
ACS/WFC observed fields of NGC 346, covering a total area of
$5\arcmin \times 5\arcmin$. North is up, and east is to the left. Image
Credits: NASA, ESA and A. Nota (STScI/ESA).\label{fig:ngc346-ima}}
\end{figure}

LHA 115-N66 or in short N66 \citep{henize56} is a very active
star-forming region in the SMC with a diameter of more than 9\arcmin\ at
a distance of about 60\,kpc \citep[e.g.][]{laney+stobie94}.
This \ion{H}{2} region is also known as DEM~S~103 \citep{davies76} 
or NGC~346 referring to the bright stellar association
located in its center. Being the largest and most luminous \ion{H}{2}
region in the SMC, NGC~346/N66 has been mapped in several wavelengths.
Observations of its area have been taken in the X-rays with {\em
XMM-Newton} and {\em Chandra} \citep{naze02, naze04}, 
in the UV with the {\em International Ultraviolet
Explorer} \citep{deboer+savage80} and the {\em
Far-Ultraviolet Spectroscopic Explorer} \citep{danforth03}, 
in the CO(2$-$1) line with the SEST telescope
\citep{contursi00}, and in the infrared with the {\em
Infrared Space Observatory} \citep{contursi00} and more
recently with the {\em Spitzer Space Telescope} \citep{bolatto07, simon07}.
The region has also been
covered by observations of the radio continuum ¸\citep{ye91, reid06} 
and of the 21-cm atomic line \citep{staveleysmith97}. 
Maps of the ionized gas have also been
made in H$\alpha$ \citep[e.g.][]{kennicutt88, lecoarer93}, 
and [{\ion{O}{3}] with NTT \citep[e.g.][]{rubio00}.

N66 is characterized by weak molecular emission and is also deficient in
\ion{H}{1}. It hosts, though, a variety of discrete H$_2$ emission peaks,
which coincide with the main features of the ionized gas, and with
compact embedded young clusters, where candidate young stellar objects
(YSOs) have been identified. The central part of the whole \ion{H}{2}
region is characterized by an oblique bright emission region, extending
from southeast to northwest, known as the bar of N66 (Fig.~\ref{fig:ngc346-ima}). 
This area is dominated by a strong UV radiation
field related to the central association NGC~346. Indeed this system
hosts the largest sample of spectroscopically confirmed OB stars in the
whole SMC \citep{massey89}, which have been the subject
of several previous investigations \citep[e.g.][]{niemela86,walborn00,evans06, hunter08}.
NGC~346 ionizes the
remaining molecular cloud to the southwest of the bar, triggering the
development of a photo-dissociated region seen as a curved front
eating into the cloud. 

Recent observations with the {\em Advanced Camera for Surveys} (ACS)
onboard the {\em Hubble Space Telescope} (HST) provided substantial
information on the stellar content of this magnificent star-forming
region by revealing a plethora of low-mass pre-main sequence (PMS)
stars. This discovery clearly indicates that NGC~346/N66 also hosts
low-mass star formation. Our photometric data set delivers
more than 98\,000 stars in total, with almost 8000 being low-mass PMS
stars \citep{gouliermis06}. The subsequent
investigation of their spatial distribution showed that these stars are
not homogeneously distributed across the region, being grouped in
discrete concentrations, apart from the association NGC~346. Most of
these ``subclusters'' are found to coincide with bright H$\alpha$
emission, and the H$_2$ emission peaks \citep{h08}.
Low-mass stars with masses \lsim~2-3~M$_\odot$ remain in their PMS phase
for $\sim$~20-30~Myr \citep[e.g.][]{sta+pal05,briceno07}. As a consequence
the rich sample of low- and intermediate-mass PMS stars, found with ACS in
NGC~346/N66, offers a unique opportunity to investigate clustered star
formation within a time-scale of the order of $\sim$~10~Myr and
length-scales of 1 to 100~pc in a detail never achieved before for a
star-forming region outside our Galaxy.

In this paper we analyze the spatial distribution of the
young PMS population unveiled by HST/ACS in the bright SMC star-forming
region NGC~346/N66 and compare the results to current theoretical
scenarios of star formation. For this purpose, we make use of our deep
ACS photometry of this region, and apply two complementary
statistical methods to identify all the compact PMS clusters and analyze
their structures. The data set used here is described in
Section~\ref{sec:data}, the statistical methods are explained in 
Section~\ref{sec:methods} and the results of our structural
analysis of the PMS stars is presented in Section~\ref{sec:analysis}.
A discussion on the star formation process revealed from
our analysis is given in Section~\ref{sec:discussion}.

\section{Description of the Data}
\label{sec:data}

The observations used in this study, taken with the Wide-Field Channel
(WFC) of ACS, and their photometry are thoroughly described by
\cite{gouliermis06}. The data were collected within
the HST GO Program 10248 and were retrieved from the HST Data
Archive\footnote{The HST Data Archive is accessible from ESO at {\tt
http://archive.eso.org/cms/hubble-space-telescope-data}, and MAST at
{\tt http://archive.stsci.edu/hst/}.}. Three overlapping images were
taken with ACS/WFC centered on the association NGC~346 in the filters
$F555W$ ($\equiv V$), $F814W$ ($\equiv I$) and $F658N$ ($\equiv$
H$\alpha$), covering an area of about $5\arcmin \times 5\arcmin$ ($\sim
80 {\rm pc} \times 80 {\rm pc}$ at the distance of the SMC). A
color-composite image of this area is shown in
Fig.~\ref{fig:ngc346-ima}. It covers the intermediate-age cluster BS~90
\citep{bica+schmitt95, rochau07},
the association NGC~346 itself, and all known components of the bright
nebula N66 \citep{henize56, davies76}.
Photometry was performed using the ACS module of the package
DOLPHOT\footnote{The ACS mode of DOLPHOT is an adaptation of the
photometry package HSTphot \citep{dolphin00}. The latest
version can be downloaded from {\tt
http://purcell.as.arizona.edu/dolphot/}.}, and after eliminating bad
detections based on the quality parameters of the detected sources
returned from DOLPHOT, more than 98\,000 stars down to $V \simeq
27.5$~mag were included in the photometric catalog. This catalog is
complete (\gsim 50\%) down to $V \approx 27$~mag. In the present study,
which deals with both low- and high-mass young populations in the area,
we make use of the upper main sequence (UMS) stars with
12~mag~\lsim~$V$~\lsim~22~mag, and the red PMS stars with $V$~\gsim~22~mag,
as they have been selected by \cite{gouliermis06}.

\section{Statistical Methods}
\label{sec:methods}

\subsection{Nearest Neighbor Density}

Star clusters are usually identified as regions of a certain overdensity with respect to
 the background stellar density.
The nearest neighbor (NN) method, introduced by \cite{casertano+hut85}
based on earlier work by \cite{vonhoerner63}, 
estimates the local source density $\rho_j$ by measuring 
the distance from each object to its $j$th nearest neighbor:
\begin{equation}
\rho_j = \frac{j - 1}{S(r_j)} m
\end{equation}
where $r_j$ is the distance of a star to its $j$th nearest neighbor, 
$S (r_j)$ the surface area with the radius $r_j$ and $m$ the average
mass of the sources ($m = 1$ when considering number densities).
The value of $j$ is chosen depending on the sample size. 
The chosen $j$~value is correlated with the sensitivity to the density
fluctuations being mapped.
A small $j$~value increases the locality 
of the density measurements at the same time as increasing sensitivity to random density 
fluctuations, whereas a large $j$ value will reduce that sensitivity at the cost of losing 
some locality.
Through the use of Monte Carlo simulations B. Ferreira \& E.~A. Lada (2008, in preparation)
find that a value of $j = 20$ is adequate to detect clusters with about 10 to 1500 members.

The positions of the cluster centers are defined as the density-weighted enhancement centers 
\citep{casertano+hut85}
\begin{equation}
x_{d,j} = \frac{\sum_i x_i \rho_j^i}{\sum_i \rho_j^i},
\end{equation}
where $x_i$ is the position vector of the $i$th cluster member and $\rho_j^i$
the $j$th NN density around this object.

Similarly, the density radius $r_d$ is defined as the density-weighted average of the distance of 
each star from the density center:
\begin{equation}
r_{d,j} = \frac{\sum_i \vert x_i - x_{d,i} \vert \rho_j^i}{\sum_i \rho_j^i}.
\label{eq:rad}
\end{equation}
It corresponds to the observational core radius \citep{casertano+hut85}.

We identify clusters as regions that have 20th NN density values $3 \sigma$ above the
average background level.

\subsection{Minimum Spanning Tree and \Q}

The second method makes use of a minimum spanning tree (MST), 
a construct from graph theory, which is defined as
the unique set of straight lines (``edges'') connecting a given set
of points without closed loops, such that the sum of the edge
lengths is a minimum \citep{boruvka26, kruskal56, prim57}.
From the MST we derive the mean edge length \lmst.
\cite{cw04} introduced the parameter $\Q = \bar{\ell}_{\rm MST}/\bar{s}$, 
which combines the normalized correlation length $\bar{s}$,
i.e.\ the mean distance between all stars, and the normalized mean edge 
length $\bar{\ell}_{\rm MST}$.
The \Q\ parameter permits to quantify the structure of a cluster and to 
distinguish between clusters with a central density concentration and 
hierarchical clusters with possible fractal substructure.
Large \Q\ values ($\Q > 0.8$) describe centrally condensed clusters
having a volume density $n(r) \propto r^{-\alpha}$, while small \Q\ values ($\Q < 0.8$) 
indicate clusters with fractal substructure.
\Q\ is correlated with the radial density exponent $\alpha$ for $\Q > 0.8$ and anticorrelated with 
the fractal dimension $D$ for $\Q < 0.8$.
The dimensionless measure \Q\ is independent of the number of objects and
of the cluster area. 
A detailed description of the method, in particular its implementation and normalization 
used for this study, is given by \cite{sk06}.

\section{Results}
\label{sec:analysis}

\subsection{The Large-Scale Structure}

\begin{figure}
\epsscale{1.2}
\plotone{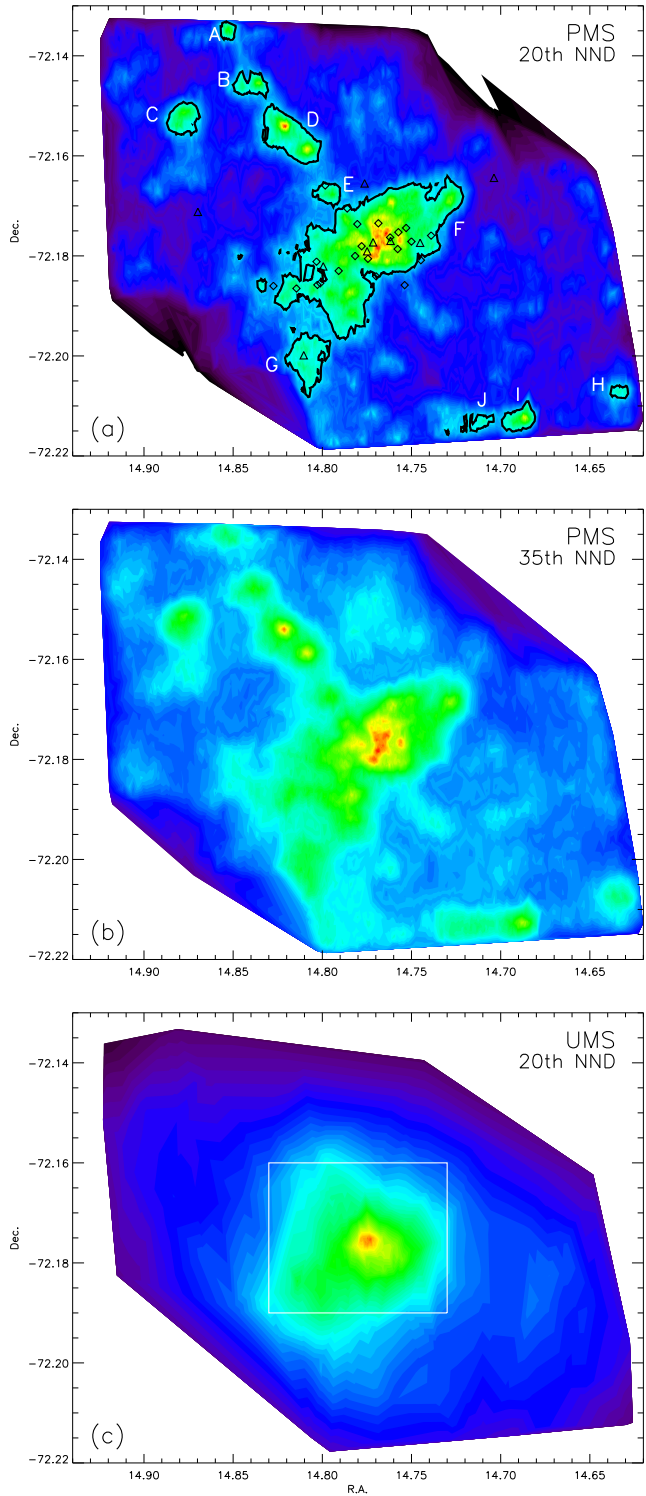}
\caption{NN density maps of NGC~346 (shown in a logarithmic scale): 
(a) 20th NN density of the PMS stars. 
The black lines indicate the cluster boundaries, defined as $3 \sigma$ above the average background density.
Also shown are the positions of the known O (diamonds) and B stars (triangles).
(b) 35th NN density of the PMS stars, emphasizing the large-scale structure.
(c) 20th NN density of the UMS stars brighter than $m_V = 18$.}
\label{fig:nnd}
\end{figure}

Figures~\ref{fig:nnd}a and \ref{fig:nnd}b show the 20th and 35th NN density maps of
the PMS stars in the NGC~346 field. While the first is used to identify clusters, the
second traces the larger-scale structure, in particular, apart from the bar of N66,
it emphasizes a filament (arm-like feature)
of enhanced stellar density extending to the north-east from the central association 
\cite[see][]{gouliermis08} and another one running in east-west direction at the 
southern edge of the observed area.

Figure~\ref{fig:nnd}c shows the 20th NN density of the UMS stars brighter than $m_V = 18$
in the same field.
While the PMS stars show a highly inhomogeneous distribution with several readily
identifiable clusters, the UMS stars show a distribution with only one central peak,
 which roughly coincides with the maximum NN density of the PMS stars.

Figure~\ref{fig:oiii} shows an image of NGC~346 in the [\ion{O}{3}] line (501.1\,nm) obtained
with the ESO NTT overlaid with the PMS NN density contours.
In the central association a high density of both gas and PMS stars is found, but the smaller
clusters north and south of the center do not coincide with a higher gas density.
The small clusters in the south appear to be aligned offset, but parallel to the gaseous filament.
When comparing the PMS stellar distribution to the 8~$\mu$m observation of the region (Figure~\ref{fig:irac4})
a remarkable correlation of the NN density peaks with the 8~$\mu$m emission peaks is found
for the central association and the clusters north of it.

\begin{figure}
\epsscale{1.15}
\plotone{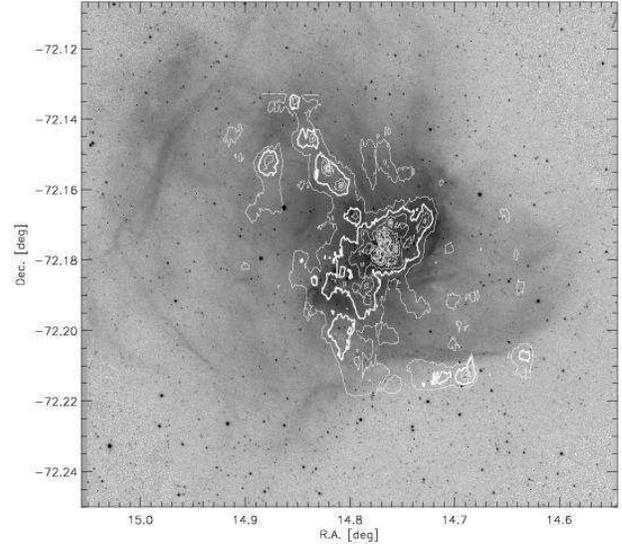}
\caption{NTT [\ion{O}{3}] (501.1\,nm) observation of NGC~346 overlaid with the PMS 20th NN density contours; the 
cluster boundary contour is shown by a thick line.
 }
\label{fig:oiii}
\end{figure}

\begin{figure}
\epsscale{1.15}
\plotone{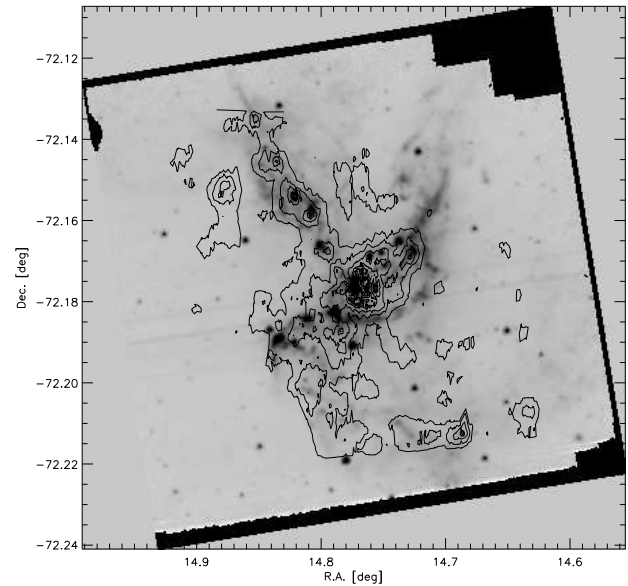}
\caption{{\em Spitzer} IRAC Channel~4 (8 $\mu$m) observation of NGC~346 overlaid with the PMS 20th NN density contours (analogous to Fig.~\ref{fig:oiii}).
 }
\label{fig:irac4}
\end{figure}

\subsection{Identified Clusters}

Using the 20th NN density and the $3 \sigma$ limit, ten PMS clusters are detected in NGC~346, labelled A to J 
from north to south (Fig.~\ref{fig:nnd}a).

The central association, Cluster F, is by far the largest cluster with 2349 PMS stars. 
Clusters A, B, D and E are parts of the filament pointing from the main association towards the north-east
more clearly visible in the 35th NN density map (Fig.~\ref{fig:nnd}b).
Cluster~C is a rather evolved open cluster with a PMS population at the age of $\sim5-15$\,Myr \citep{h08},
and it may be not related to the much younger other clusters.
While Clusters~B and D show two clearly visible separate density peaks, they are picked up as single clusters
using the $3 \sigma$ limit in the NN density maps.
However, when using a  $4 \sigma$ threshold, cluster D breaks up into two clusters (D1 and D2),
while of cluster~B only the western peak remains (B2).
Clusters~C, D, and E are also identified by \cite{h08} 
as their Clusters 1, 2 and 3. Their Clusters 4 and 5 are part of the main association 
in the NN method.
Clusters C, D, and E correspond to the subclusters Sc~16, Sc~14 + Sc~15, and Sc~13 identified by \cite{sabbi07},
while the central association is broken into eleven subclusters (Sc~1 to Sc~11) by these authors.
It is not possible to reproduce the number and locations of the central subclusters of \cite{sabbi07}
using the NN method even with different $j$ values ($5 \le j \le 50$) and different threshold levels 
($2 \sigma \le \rho_{\rm j} \le 6 \sigma$).

Table~\ref{tab:clusters} lists the identified PMS clusters with the positions of their 20th NN density
centers (columns~2 and 3), the number of cluster members (column~4), their density radius (Eq.~\ref{eq:rad}) in pc 
(assuming a distance of 62\,kpc; column~5), maximum 20th NN density (column~6), and \Q\ (column~7).

%
%

\begin{table}[t]
 \centering
  \caption{Detected PMS clusters}
  \label{tab:clusters}
  \begin{tabular}{l c c r c r c l}
  \hline
Cluster & RA (J2000)   &      Dec (J2000)  &  $n_*$ & $r_{\rm d}$ & $\rho_{20}^{\rm max}$ & \Q  \\

& (deg) &     (deg) &   & (pc)    & (pc$^{-2}$)   &  & \\
\hline
A & 14.85263  & $-$72.13482  &   27 & 0.57 &  5.35  &   0.93  \\
B & 14.83835  & $-$72.14563  &   59 & 1.46 &  8.38  &   0.59  \\
C & 14.87806  & $-$72.15229  &   90 & 1.79 &  6.11  &   0.74  \\ 
D & 14.81802  & $-$72.15552  &  257 & 2.92 & 36.35  &   0.65  \\
E & 14.79577  & $-$72.16745  &   47 & 1.21 &  5.34  &   0.92  \\
F & 14.76906  & $-$72.17724  & 2349 & 5.26 & 40.02  &   0.68  \\
G & 14.80844  & $-$72.20081  &  139 & 3.14 &  3.45  &   0.70  \\
H & 14.63330  & $-$72.20700  &   18 & 0.70 &  2.49  &   0.86  \\
I & 14.68797  & $-$72.21261  &   74 & 1.10 & 13.22  &   0.80  \\
J & 14.71087  & $-$72.21309  &   26 & 1.31 &  2.83  &   0.78  \\
\hline
\end{tabular}
\end{table}

\subsection{Clustering Parameters}

\begin{figure}
\epsscale{1.0}
\plotone{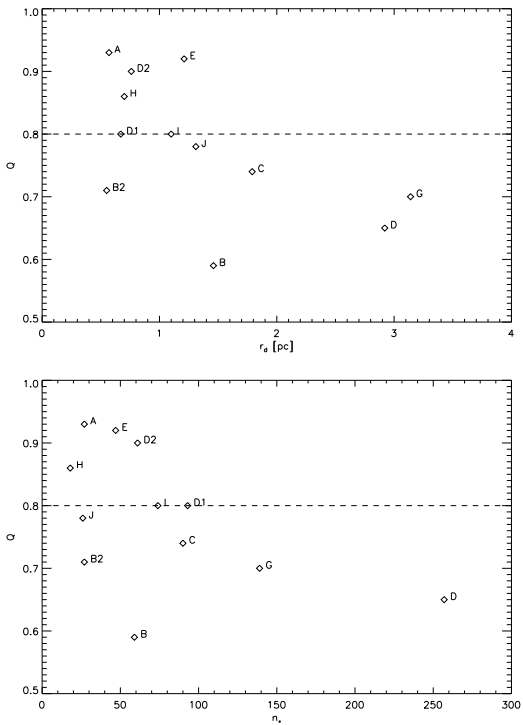}
\caption{\Q\ versus density radius (upper panel) and number of stars
(lower panel) for the identified PMS clusters (cluster F not shown).
The dashed line at $\Q = 0.8$ indicates the division between hierarchical and
centrally condensed clusters.}
\label{fig:Q_vs_size}
\end{figure}

The detected clusters show a wide variety in the \Q\ parameter, ranging from strongly centrally
condensed clusters with $\Q > 0.9$ (clusters A, E) to highly hierarchical clusters with
multiple density peaks and values $\Q < 0.7$ (clusters B, D).
Also the central association (F) with its several density peaks shows, as expected, a hierarchical 
distribution ($\Q = 0.68$). However, this value has to be taken with care, since due to a degeneracy
of the \Q\ parameter, highly elongated structures like this one systematically show lower \Q\ values
\citep{bastian08}. Applying the correction suggested by \citet{bastian08} changes the value
to $\Q \approx 0.74$, still in the hierarchical regime.
To avoid these uncertainties, and to compare the values for PMS and UMS stars,
we use a quadrilateral area around the central region with $14\fdg73 \le \alpha \le 14\fdg83$ and 
$-72\fdg19 \le \delta \le -72\fdg16$ (shown as a white box in Fig.~\ref{fig:nnd}c).
Both PMS and UMS stars in this region are equally centrally concentrated ($\Q = 0.81$). 
As far as the entire NGC~346 area is concerned, the UMS stars show the same \Q\ value,
while the PMS stars show a hierarchical distribution ($\Q = 0.73$).

The subclusters of multi-peak clusters can be centrally concentrated or again hierarchical.
The two subclusters D1 and D2 show \Q\ values of 0.80 and 0.90, respectively, while B2
has a value of $\Q = 0.71$.

The \Q\ parameters show a moderate correlation with the size of the cluster measured by $r_{\rm d}$
and the number of stars, illustrated in Fig.~\ref{fig:Q_vs_size}. 
This is in remarkable contradiction to the embedded clusters found in nearby Galactic molecular clouds,
where smaller clusters tend to have \Q\ values well in the hierarchical regime \citep{skf08}.
This may be due to the fact that we are only tracing the more evolved PMS stars in NGC~346 
that probably show a more centrally condensed structure than Class~0/1 protostars, which are expected
to be hierarchically clustered \citep{skf08}.

\subsection{\Q\ versus Magnitude}

%

We investigate the variation of \Q\ as a function of magnitude (and thus, mass) for the PMS and UMS sample.
This is done for the entire area as well as for the central region.
\Q\ was calculated for sources brighter than a certain magnitude
in steps of 0.5\,mag as long as the sample contained at least 40 objects.
The resulting \Q\ versus magnitude plots for PMS stars and UMS stars are shown in 
Fig.~\ref{fig:Q_mag_pms} and \ref{fig:Q_mag_ums}, respectively.
Both PMS and UMS stars show a decline of \Q\ with fainter magnitudes when looking
at the entire area, indicating that the massive stars are more centrally condensed
than lower-mass objects. 
However, both samples show no correlation of \Q\ with magnitude when considering
the central region.

\begin{figure}
\epsscale{1.1}
\plotone{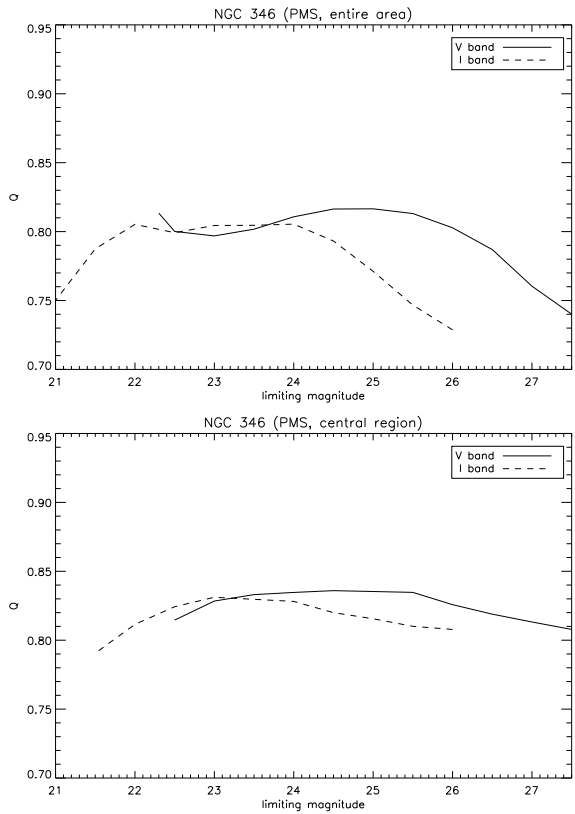}
\caption{\Q\ versus limiting magnitude for the PMS stars in the entire area
(upper panel) and in the central region (lower panel).}
\label{fig:Q_mag_pms}
\end{figure}

\begin{figure}
\epsscale{1.1}
\plotone{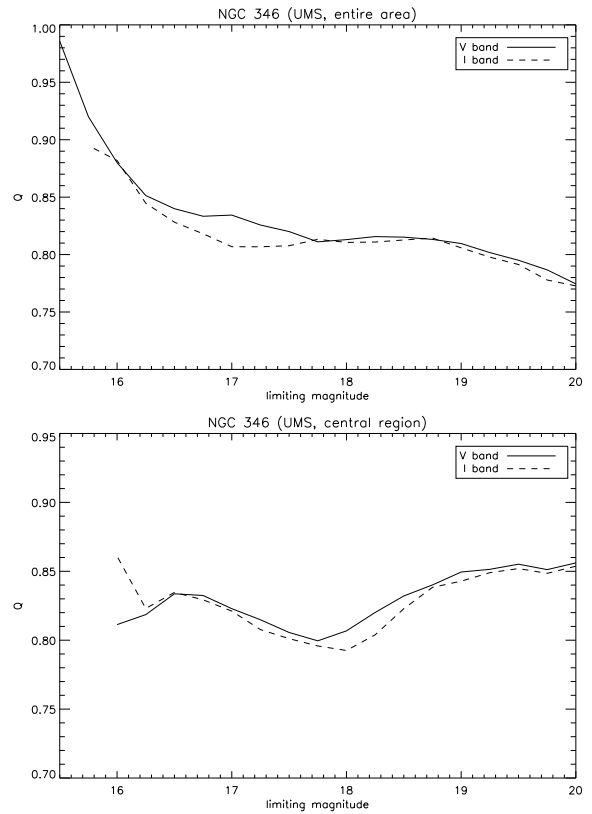}
\caption{\Q\ versus limiting magnitude for the UMS stars in the entire area
(upper panel) and in the central region (lower panel).}
\label{fig:Q_mag_ums}
\end{figure}

\subsection{OB Stars}

In Fig.~\ref{fig:nnd}a the positions of the spectroscopically verified
OB~stars in NGC~346 by \cite{massey89} are shown overlaid on the
20th NN density map. While it is apparent that the OB~stars are
centrally concentrated in the association, it is interesting to note
that they are not located in the densest (yellow-red) regions of PMS
stars. This indicates that the density peaks of PMS stars observed in
Cluster~F may be an observational bias, due to the fact that the OB~stars
outshine the faint PMS stars in their vicinity. This provides further
support to our identification of the association as a single, not
sub-clustered system from both observed UMS and PMS populations. As far
as the clustering behavior of the OB stars alone is concerned, they do
not seem to form any central condensation. Indeed, when considering the
27 OB~stars lying in the association, they show a distribution in the
hierarchical range with $\Q = 0.77$.

\section{Discussion and Conclusions}
\label{sec:discussion}

\subsection{Theoretical Concepts}

With the advancement of infrared detectors and the ability to do wide-area surveys, it has been recognized that a large fraction of all stars in the Milky Way form in clusters and aggregates of various size and mass scales and that isolated or widely distributed star formation is the exception rather than the rule \citep{lada03}. 
It is therefore highly interesting to study the clustering behavior in neighboring galaxies as well.
The complex hierarchical structure of molecular clouds provides a natural explanation for clustered star formation. 
Molecular clouds vary enormously in size and mass. In small, low-density, clouds stars form with low efficiency, more or less in isolation or scattered around in small groups of up to a few dozen members. Denser and more massive clouds may build up stars in associations and clusters of a few hundred members.  This appears to be the most common mode of star formation, at least in the solar neighborhood \citep{adams01}. The formation of dense rich clusters with thousands of stars or more is rare.  The cluster NGC 346, therefore, is a very extreme case of clustered star formation. 

Star formation is governed by the complex interplay between gravitational attraction in molecular clouds and any opposing processes such as supersonic turbulence, thermal pressure and magnetic fields. 

There are two main competing models that describe the evolution of the cloud cores.  It was proposed in the 1980's that cores in low-mass star-forming regions evolve quasi-statically in magnetically subcritical clouds \citep{shu87}. Gravitational contraction is mediated by ambipolar diffusion \citep{Mou76, Mou91} causing a redistribution of magnetic flux until the inner regions of the core become supercritical and go into dynamical collapse.  Measurements of the Zeeman splitting of molecular lines in nearby cloud cores, however, indicate magnetic field strengths that fall below the critical value, in some cases only by a small margin, in many cases however, by factors of many \citep{crutcher99, bourke01,crutcher08}. Magnetic fields 
therefore seem not to be the dominant agent regulating stellar birth. 

This has lead to the suggestion that the supersonic turbulence ubiquitously observed in molecular clouds is the main physical process governing star formation \citep{maclow+klessen04,ballesteros07,mckee+ostriker07}. Supersonic turbulence plays a dual role. On large scales it can support clouds against contraction, however, on small scales it can provoke localized collapse. Turbulence establishes a complex network of interacting shocks, where dense cores form at the stagnation points of convergent flows. The density can be large enough for gravitational collapse to set in. However, the fluctuations in turbulent velocity fields are highly transient.  The random flow that creates local density enhancements can disperse them again.  For local collapse to actually result in the formation of stars, high density fluctuations must collapse on time scales shorter than the typical time interval between two successive shock passages.  Only then are they able to `decouple' from the ambient flow and survive subsequent shock interactions.  The shorter the time between shock passages, the less likely these fluctuations are to survive. Hence, the timescale and efficiency of protostellar core formation depend strongly on the wavelength and strength of the driving source \citep{klessen00b,Heitschetal2001,Vazquez03}, and accretion histories of individual protostars are strongly time varying \citep{Klessen2001b,sk04}.

Interstellar turbulence is observed to be dominated by large-scale velocity modes \citep{maclow00b,Ossenkopfetal2001,Ossenkopf02}. This implies it is very efficient  in sweeping up molecular cloud material, creating massive coherent structures which can go into large-scale gravitational collapse. In the extreme case, dense clusters with many thousand stars build up on one or two crossing times \citep{Hartmann01,elmegreen00,elmegreen07}. Numerical simulations of gravoturbulent cloud fragmentation indicate that in such cases star clusters would build up in a hierarchical fashion with sub-clusters which evolve dynamically and merge together to build up the final, centrally-condensed cluster \citep{klessen00,clarke00,bonnell03,clark05b,BonnellBate2006},
with stellar feedback strongly influencing this process \citep[][]{krumholz07}.
\citet{sk06} investigated the temporal evolution of a large
set of numerical simulations of gravoturbulent fragmentation and showed that the
\Q\ parameter of all resulting clusters increases with time.
This is confirmed by observations of embedded clusters in nearby
molecular clouds, where the younger Class 0/1 sources
show substantially lower \Q\ values than the more evolved Class 2/3 objects
\citep{skf08}.

\subsection{The Structure of NGC 346/N66}

Within our investigation of the spatial distribution of PMS stars in the
NGC~346/N66 region using the nearest neighbor and minimum spanning tree
methods an increased NN density of PMS stars is found in the central
association and in two filamentary structures, one extending to the
north-east from the center of N66, and another at the southern edge of the observed
area, extending from east to west.
We identify ten individual clusters of PMS stars in the region, with the central
association being the largest. While several PMS density peaks are found 
within the association NGC~346, the system could not be resolved
into individual subclusters. Our methods show evidence of centrally
concentrated UMS stars and PMS stars in a hierarchical distribution. The
PMS density peaks of the association coincide with peaks of both [{\ion{O}{3}]
and 8~$\mu$m emission, indicating that the system is in an early
evolutionary stage. 

The identified PMS clusters in the whole area of NGC~346/N66 show a wide
range of morphologies from hierarchical multi-peak configurations to
centrally condensed clusters. Only about 40\% of the PMS stars in the
investigated area belong to the ten identified clusters. The remaining 60\% of
PMS stars is found concentrated in small (statistically insignificant)
groups and mostly dispersed over the entire area. 
This fraction of
clustered PMS stars is comparable to that found in
Galactic star-forming regions. 
In their SCUBA survey of the Perseus molecular cloud \cite{hatchell05} 
found that 40~-~60\% of the protostars and
prestellar cores are located in small clusters ($<$~50~M$_\odot$) and
isolated objects. Investigations in other Galactic star-forming regions
 like Ophiuchus, Orion and Monoceros yield a percentage of
distributed populations between 20 and 60\% with a ``typical'' value of
25\% \citep{carpenter00, allen07}. However, there are
uncertainties due to the incompleteness of the samples and the definition
of the clusters. Moreover, the clustering properties themselves depend on
the age of the population considered and on the masses of the stars. As a
consequence, it is very difficult to compare and interpret these
numbers.

Based on the analysis of \Q\ versus
magnitude, both PMS and UMS stars show roughly the same degree of
central concentration over the entire magnitude range. When we consider
the entire area, \Q\ is found to decrease with magnitude, providing
evidence that both PMS and UMS stars are centrally concentrated in
respect to the whole observed region of NGC~346. This, especially for
the bright stars, is expected considering that the central association
contains the largest fraction of stars. However, when we consider only
the central region of the area, \Q\ does not seem to depend strongly on
magnitude, and therefore no evidence of mass segregation in the central
association is found.

Whether the clusters in NGC~346 have formed coevally
\citep{sabbi07} or in a sequential process over several Myr
\citep{massey89, rubio00} is still under debate.
As young clusters seem to evolve from a hierarchical
configuration to a centrally concentrated one with time,
the fact that the identified clusters in NGC~346 show a wide range in
their structures with \Q\ values ranging from 0.65 to 0.93, may indeed
indicate that they are at different evolutionary stages. However, the
absolute \Q\ values are not correlated with the cluster age, but
probably depend on a variety of additional factors like initial conditions 
or turbulent energy in the cloud \citep{skf08}, and therefore
this is no direct proof that the clusters have formed at different times.
However, the rather high \Q\ values of the southern clusters (H, I, J) around or above 0.8
agree with other age indicators.
While the peaks of the central concentration and the clusters
to the north of it coincide with the peaks of dust emission (indicated by
the  8~$\mu$m observations) and gas (traced by [{\ion{O}{3}}]),
the southern clusters appear to be associated with less amounts
of gas and dust, suggesting an older age of these clusters.
\citet{h08} find that the northern clusters comprise a younger
PMS population (not older than 2.5\,Myr) than the other clusters.
Recently, \citet{gouliermis08} presented evidence that these clusters 
are the product of triggered star formation.
In combination, all these indications strongly suggest that the identified PMS clusters are
not the product of a single star formation event.

In conclusion our investigation of the spatial distribution of PMS stars
in the region of NGC~346/N66 using the nearest neighbor and minimum
spanning tree methods showed that this region appears to be the product
of hierarchical star formation, where the individual clusters show
different structures according to their size, evolutionary stage,
position in the cloud and possible additional internal or environmental
factors.
The overall structure and statistical properties of the region are
consistent with numerical calculations of gravoturbulent cloud
fragmentation \citep[e.g.][]{klessen00b,clark05a,BonnellBate2006}.
NGC~346/N66 is a unique case, where the complexity of star
formation can be observed in its entire length-scale.

\acknowledgments
SS and DAG are supported by the {\em Deutsche Forschungs\-gemeinschaft} (DFG) through individual grants
SCHM 2490/1-1 and GO 1659/1-1, respectively.
RSK acknowledges partial support from the DFG via grant KL 1358/1-3 and via the priority program SFB~439 {\em Galaxies in the Young Universe}.
This work is based on observations made with the NASA/ESA {\em Hubble Space Telescope},
obtained from the data archive at the Space Telescope Science Institute (STScI).
STScI is operated by the Association of Universities for Research in Astronomy, Inc.\ under NASA contract 
NAS 5-26555.
This work also makes use of observations made with the ESO NTT at La Silla Observatory under program ID 56.C-0379
and observations made with the {\it Spitzer Space Telescope}, which is operated by the Jet Propulsion Laboratory,
California Institute of Technology, under a contract with NASA.



\begin{thebibliography}{}

\bibitem[Adams \& Myers(2001)]{adams01}
Adams, F.~C., \& Myers, P.~C.\ 2001, \apj, 553, 744

\bibitem[Allen et al.(2007)]{allen07}
Allen, L., et al.\ 2007, in Protostars and Planets V, 
ed.\ B. Reipurth, D. Jewitt, \& K. Keil (Tucson: Univ.\ Arizona Press), 361

\bibitem[Ballesteros-Paredes et al.(2007)]{ballesteros07}
Ballesteros-Paredes, J., Klessen, R.~S., Mac Low, M.-M., \& V\'{a}zquez-Semadeni, E. 2007,
in Protostars and Planets V, ed.\ B. Reipurth, D. Jewitt, \& K. Keil (Tucson: Univ.\ Arizona Press), 63

\bibitem[Bastian et al.(2008)]{bastian08}
Bastian, N., Gieles, M., Ercolano, B., \& Gutermuth, R.\ 2008, \mnras, in press (arXiv:0810.3190)

\bibitem[Bica \& Schmitt(1995)]{bica+schmitt95}
Bica, E.~L.~D., \& Schmitt, H.~R. 1995, \apjs, 101, 41

\bibitem[Bolatto et al.(2007)]{bolatto07} 
Bolatto, A., et al. 2007, \apj, 655, 212

\bibitem[Bonnell \& Bate(2006)]{BonnellBate2006}
Bonnell, I.~A., \& Bate, M.~R.\ 2006, \mnras, 370, 488 

\bibitem[Bonnell et al.(2003)]{bonnell03}
Bonnell, I.~A., Bate, M.~R., \& Vine, S.~G.\ 2003, \mnras, 343, 413 

\bibitem[Bor\r{u}vka(1926)]{boruvka26}
Bor\r{u}vka, O. 1926, Pr{\'a}ce moravsk\'e p\v{r}\'{\i}rodov\v{e}deck\'e spole\v{c}nosti, 3, 37

\bibitem[Bourke et al.(2001)]{bourke01}
Bourke, T.~L., Myers, P.~C., Robinson, G., \& Hyland, A.~R.\ 2001, \apj, 554, 916 

\bibitem[Brice\~{n}o et al.(2007)]{briceno07}
Brice{\~n}o, C., Preibisch, T., Sherry, W.~H., Mamajek, E.~A., Mathieu, R.~D., Walter, 
F.~M., \& Zinnecker, H. 2007,
in Protostars and Planets V, ed.\ B. Reipurth, D. Jewitt, \& K. Keil (Tucson: Univ.\ Arizona Press), 345

\bibitem[Carpenter(2000)]{carpenter00}
Carpenter, J.~M.\ 2000, \aj, 120, 3139

\bibitem[Cartwright \& Whitworth(2004)]{cw04}
Cartwright, A., \& Whitworth, A.~P. 2004, \mnras, 348, 589

\bibitem[Casertano \& Hut(1985)]{casertano+hut85}
Casertano, S., \& Hut, P. 1985, \apj, 298, 80

\bibitem[Clark \& Bonnell(2005)]{clark05b}
Clark, P.~C., \& Bonnell, I.~A.\ 2005, \mnras, 361, 2 

\bibitem[Clark et al.(2005)]{clark05a}
Clark, P.~C., Bonnell, I.~A., Zinnecker, H., \& Bate, M.~R.\ 2005, \mnras, 359, 809

\bibitem[Clarke et al.(2000)]{clarke00}
Clarke, C.~J., Bonnell, I.~A., \& Hillenbrand, L.~A.\ 2000, in Protostars and Planets IV, 
ed.\ V. Mannings, A.~P. Boss, \& S.~S. Russell (Tucson: Univ.\ Arizona Press), 151

\bibitem[Contursi et al.(2000)]{contursi00}
Contursi, A., et al. 2000, \aap, 362, 310

\bibitem[Crutcher(1999)]{crutcher99}
Crutcher, R.~M.\ 1999, \apj, 520, 706 

\bibitem[Crutcher et al.(2008)]{crutcher08}
Crutcher, R.~M., Hakobian, N., \& Troland, T.~H.\ 2008, ApJ, accepted (arXiv:0807.2862)

\bibitem[Danforth et al.(2003)]{danforth03}
Danforth, C.~W., Sankrit, R., Blair, W.~P., Howk, J.~C., \& Chu, Y.-H.
2003, \apj, 586, 1179

\bibitem[Davies et al.(1976)]{davies76} 
Davies, R.~D., Elliott, K.~H., \& Meaburn, J. 1976, \memras, 81, 89

\bibitem[de Boer \& Savage(1980)]{deboer+savage80}
de Boer, K.~S., \& Savage, B.~D. 1980, \apj, 238, 86

\bibitem[Dolphin(2000)]{dolphin00}
Dolphin, A. E. 2000, \pasp, 112, 1383

\bibitem[Efremov \& Elmegreen(1998)]{ef+el98}
Efremov, Y.~N., \& Elmegreen, B.~G. 1998, \mnras, 299, 588

\bibitem[Elmegreen(2000)]{elmegreen00}
Elmegreen, B.~G.\ 2000, \apj, 530, 277 

\bibitem[Elmegreen(2006)]{elm06}
Elmegreen, B.~G. 2006, in Globular Clusters, Guide to Galaxies, ed.\ T.
Richtler et al.\ (Berlin: ESO/Springer), in press
(astro-ph/0605519)

\bibitem[Elmegreen(2007)]{elmegreen07}
Elmegreen, B.~G.\ 2007, \apj, 668, 1064

\bibitem[Elmegreen et al.(2000)]{elm00}
Elmegreen, B.~G., Efremov, Y., Pudritz, R.~E., \& Zinnecker H. 2000,
in Protostars and Planets IV, ed.\ V. Mannings, A.~P. Boss, \& S.~S. Russell 
(Tucson: Univ.\ Arizona Press), 179

\bibitem[Evans et al.(2006)]{evans06}
Evans, C.~J., Lennon, D.~J., Smartt, S.~J., \& Trundle, C.\ 2006, \aap, 456, 623

\bibitem[Gouliermis et al.(2006)]{gouliermis06}
Gouliermis, D.~A., Dolphin, A.~E., Brandner, W., \& Henning, T.\ 2006,
\apjs, 166, 549

\bibitem[Gouliermis et al.(2008)]{gouliermis08}
Gouliermis, D.~A., Chu, Y.-H., Henning, T., Brandner, W., Gruendl, R.~A., Hennekemper, E., \& Hormuth, F.\ 2008,
\apj, 688, 1050

\bibitem[Hartmann et al.(2001)]{Hartmann01}
Hartmann, L., Ballesteros-Paredes, J., \& Bergin, E.~A.\ 2001, \apj, 562, 852 

\bibitem[Hatchell et al.(2005)]{hatchell05}
Hatchell, J., Richer, J.~S., Fuller, G.~A., Qualtrough, C.~J., Ladd, E.~F., \& Chandler, C.~J.\ 2005, \aap, 440, 151

\bibitem[Henize(1956)]{henize56}
Henize, K. G. 1956, \apjs, 2, 315

\bibitem[Hennekemper et al.(2008)]{h08} 
Hennekemper, E., Gouliermis, D.~A., Henning, T., Brandner, W., \& Dolphin, A.~E. 2008, \apj, 672, 914

\bibitem[Heitsch et al.(2001)]{Heitschetal2001}
Heitsch, F., Mac Low, M.-M., \& Klessen, R.~S.\ 2001, \apj, 547, 280 

\bibitem[Hunter et al.(2008)]{hunter08}
Hunter, I., et al.\ 2008, \aap, 479, 541

\bibitem[Kennicutt(1988)]{kennicutt88}
Kennicutt, R.~C., Jr.\ 1988, \apj, 334, 144

\bibitem[Klessen(2001)]{Klessen2001b}
Klessen, R.~S.\ 2001, \apjl, 550, L77 

\bibitem[Klessen \& Burkert(2000)]{klessen00}
Klessen, R.~S., \& Burkert, A.\ 2000, \apjs, 128, 287

\bibitem[Klessen et al.(2000)]{klessen00b}
Klessen, R. S., Heitsch, F., \& Mac Low, M.-M. 2000, \apj, 535, 887

\bibitem[Krumholz et al.(2007)]{krumholz07}
Krumholz, M.~R., Klein, R.~I., \& McKee, C.~F.\ 2007, \apj, 656, 959

\bibitem[Kruskal(1956)]{kruskal56}
Kruskal, J.~B. Jr. 1956, Proc.\ Amer.\ Math.\ Soc., 7, 48

\bibitem[Lada \& Lada(2003)]{lada03}
Lada, C.~J., \& Lada, E.~A.\ 2003, \araa, 41, 57

\bibitem[Laney \& Stobie(1994)]{laney+stobie94}
Laney, C.~D., \& Stobie, R.~S. 1994, \mnras, 266, 441 

\bibitem[Le Coarer et al.(1993)]{lecoarer93}
Le Coarer, E., Rosado, M., Georgelin, Y.~P., Viale, A., \& Goldes, G., 1993, \aap, 280, 365

\bibitem[Mac Low \& Ossenkopf(2000)]{maclow00b}
Mac Low, M.-M., \& Ossenkopf, V.\ 2000, \aap, 353, 339 

\bibitem[Mac Low \& Klessen(2004)]{maclow+klessen04}
Mac Low, M.-M., \& Klessen, R.~S. 2004, Rev. Mod. Phys., 76, 125

\bibitem[Massey et al.(1989)]{massey89}
Massey, P., Parker, J.~W., \& Garmany, C.~D. 1989, \aj, 98, 1305

\bibitem[McKee \& Ostriker(2007)]{mckee+ostriker07}
McKee, C. F., \& Ostriker, E.~C. 2007, \araa, 45, 565

\bibitem[Mouschovias(1991)]{Mou91}
Mouschovias, T.~C.\ 1991, NATO ASIC Proc.~342: The Physics of Star Formation and Early Stellar 
Evolution, 61 

\bibitem[Mouschovias \& Spitzer(1976)]{Mou76}
Mouschovias, T.~C., \& Spitzer, L., Jr.\ 1976, \apj, 210, 326 

\bibitem[Naz\'{e} et al.(2002)]{naze02}
Naz\'{e}, Y., et al. 2002, \apj, 580, 225

\bibitem[Naz\'{e} et al.(2004)]{naze04} 
Naz\'{e}, Y., Manfroid, J., Stevens, I. R., Corcoran, M. F., \& Flores, A.
2004, \apj, 608, 208

\bibitem[Niemela et al.(1986)]{niemela86}
Niemela, V. S., Marraco, H. G., \& Cabanne, M. L. 1986, \pasp, 98, 1133

\bibitem[Ossenkopf \& Mac Low(2002)]{Ossenkopf02}
Ossenkopf, V., \& Mac Low, M.-M.\ 2002, \aap, 390, 307 

\bibitem[Ossenkopf et al.(2001)]{Ossenkopfetal2001}
Ossenkopf, V., Klessen, R.~S., \& Heitsch, F.\ 2001, \aap, 379, 1005 

\bibitem[Prim(1957)]{prim57}
Prim, R.~C. 1957, Bell Syst.\ Tech.\ J., 36, 1389

\bibitem[Reid et al.(2006)]{reid06}
Reid, W.~A., et al.\ 2006, \mnras, 367, 1379

\bibitem[Rochau et al.(2007)]{rochau07}
Rochau, B., Gouliermis, D.~A., Brandner, W., Dolphin, A.~E., \& Henning,
T. 2007, \apj, 664, 322

\bibitem[Rubio et al.(2000)]{rubio00}
Rubio, M., et al. 2000, \aap, 359, 1139

\bibitem[Sabbi et al.(2007)]{sabbi07}
Sabbi, E., et al.\ 2007, \aj, 133, 44 

\bibitem[Scalo(1985)]{scalo85}
Scalo, J.~M. 1985, in Protostars and Planets II, ed.\ D.~C. Black
\& M.~S. Matthews (Tucson: Univ.\ Arizona Press), 201

\bibitem[Schmeja \& Klessen(2004)]{sk04}
Schmeja, S., \& Klessen, R.~S. 2004, \aap, 419, 405 

\bibitem[Schmeja \& Klessen(2006)]{sk06}
Schmeja, S., \& Klessen, R.~S. 2006, \aap, 449, 151

\bibitem[Schmeja et al.(2008)]{skf08}
Schmeja, S., Kumar, M.~S.~N., \& Ferreira, B. 2008, \mnras, 389, 1209

\bibitem[Shu et al.(1987)]{shu87}
Shu, F.~H., Adams, F.~C., \& Lizano, S.\ 1987, \araa, 25, 23 

\bibitem[Simon et al.(2007)]{simon07}
Simon, J. D., et al. 2007, \apj, 669, 327

\bibitem[Stahler \& Palla(2005)]{sta+pal05}
Stahler, S. W., \& Palla, F. 2005, The Formation of Stars (Weinheim: Wiley-VCH)

\bibitem[Staveley-Smith et al.(1997)]{staveleysmith97}
Staveley-Smith, L., Sault, R.~J., Hatzidimitriou, D., Kesteven, M.~J., \& 
McConnelle, D. 1997, \mnras, 289, 225

\bibitem[V\'{a}zquez-Semadeni(2004)]{vazquez04}
V\'{a}zquez-Semadeni, E. 2004, \apss, 292, 187

\bibitem[V{\'a}zquez-Semadeni et al.(2003)]{Vazquez03}
V{\'a}zquez-Semadeni, E., Ballesteros-Paredes, J., \& Klessen, R.~S.\ 2003, \apjl, 585, L131 

\bibitem[von Hoerner(1963)]{vonhoerner63}
von Hoerner, S. 1963, \zap, 57, 47

\bibitem[Walborn et al.(2000)]{walborn00}
Walborn, N.~R., et al. 2000, \pasp, 112, 1243

\bibitem[Ye et al.(1991)]{ye91}
Ye, T., Turtle, A.~J., \& Kennicutt, R.~C. Jr. 1991, \mnras, 249, 722

\end{thebibliography}
\end{document}